\documentstyle[12pt]{article}
\def\beq{\begin{equation}}
\def\eeq{\end{equation}}

\def\as{\alpha_s}

\def\xp{\Xi_c^+}
\def\xn{\Xi_c^0}

\def\ap#1#2#3 {Ann. Phys. (NY) {\bf#1} (19#2) #3}
\def\apj#1#2#3 {Astrophys. J. {\bf#1} (19#2) #3}
\def\apjl#1#2#3 {Astrophys. J. Lett. {\bf#1} (19#2) #3}
\def\app#1#2#3 {Acta. Phys. Pol. {\bf#1} (19#2) #3}
\def\ar#1#2#3 {Ann. Rev. Nucl. Part. Sci. {\bf#1} (19#2) #3}
\def\cpc#1#2#3 {Computer Phys. Comm. {\bf#1} (19#2) #3}
\def\err#1#2#3 {{\it Erratum} {\bf#1} (19#2) #3}
\def\ib#1#2#3 {{\it ibid.} {\bf#1} (19#2) #3}
\def\jmp#1#2#3 {J. Math. Phys. {\bf#1} (19#2) #3}
\def\ijmp#1#2#3 {Int. J. Mod. Phys. {\bf#1} (19#2) #3}
\def\jetp#1#2#3 {JETP Lett. {\bf#1} (19#2) #3}
\def\jpg#1#2#3 {J. Phys. G. {\bf#1} (19#2) #3}
\def\mpl#1#2#3 {Mod. Phys. Lett. {\bf#1} (19#2) #3}
\def\nat#1#2#3 {Nature (London) {\bf#1} (19#2) #3}
\def\nc#1#2#3 {Nuovo Cim. {\bf#1} (19#2) #3}
\def\nim#1#2#3 {Nucl. Instr. Meth. {\bf#1} (19#2) #3}
\def\np#1#2#3 {Nucl. Phys. {\bf#1} (19#2) #3}
\def\pcps#1#2#3 {Proc. Cam. Phil. Soc. {\bf#1} (#2) #3}
\def\pl#1#2#3 {Phys. Lett. {\bf#1} (19#2) #3}
\def\prep#1#2#3 {Phys. Rep. {\bf#1} (19#2) #3}
\def\prev#1#2#3 {Phys. Rev. {\bf#1} (19#2) #3}
\def\prl#1#2#3 {Phys. Rev. Lett. {\bf#1} (19#2) #3}
\def\prs#1#2#3 {Proc. Roy. Soc. {\bf#1} (19#2) #3}
\def\ptp#1#2#3 {Prog. Th. Phys. {\bf#1} (19#2) #3}
\def\ps#1#2#3 {Physica Scripta {\bf#1} (19#2) #3}
\def\rmp#1#2#3 {Rev. Mod. Phys. {\bf#1} (19#2) #3}
\def\rpp#1#2#3 {Rep. Prog. Phys. {\bf#1} (19#2) #3}
\def\sjnp#1#2#3 {Sov. J. Nucl. Phys. {\bf#1} (19#2) #3}
\def\spj#1#2#3 {Sov. Phys. JEPT {\bf#1} (19#2) #3}
\def\spu#1#2#3 {Sov. Phys. Usp. {\bf#1} (19#2) #3}
\def\zp#1#2#3 {Zeit. Phys. {\bf#1} (19#2) #3} 

\begin{document}
\begin{titlepage}
\begin{center}
{\Large \bf Theoretical Physics Institute \\
University of Minnesota \\}  \end{center}
\vspace{0.3in}
\begin{flushright}
TPI-MINN-96/4-T \\
UMN-TH-1425-96 \\
April 1996
\end{flushright}
\vspace{0.4in}
\begin{center}
{\Large \bf  Spectator effects in semileptonic decay of charmed baryons\\}
\vspace{0.2in} 
{\bf M.B. Voloshin  \\ } 
Theoretical Physics Institute, University of Minnesota, Minneapolis, MN 
55455 \\ and \\ 
Institute of Theoretical and Experimental Physics, Moscow, 117259 \\[0.2in]

{\bf   Abstract  \\ }
\end{center}

It is shown that the absolute rate of the semileptonic decays of the $\Xi_c$ 
baryons can be as much as two times higher than the semileptonic decay rate 
of $\Lambda_c$ due to the Pauli interference of the strange quark.  This 
interference effect is still larger in the semileptonic decay rate of 
$\Omega_c$. An experimental measurement of the difference of these rates 
would thus provide an important piece of information on the spectator 
effects in decays of hadrons containing a heavy quark.  
\end{titlepage} 

\section{Introduction}
The differences of the lifetimes of hadrons containing a heavy quark are 
attracting considerable interest ever since the early experimental 
evidence$^{\cite{delco}}$ for unequal lifetimes of the charmed $D^\pm$ and 
$D^0$ mesons. Presently measured lifetimes of weakly decaying charmed 
hadrons span more than one order of magnitude:  $\tau(D^\pm)/\tau(\Omega_c) 
\sim 20$. For the hadrons, containing the $b$ quark, the relative 
differences in lifetimes are much smaller and the experimental situation 
with their precise values is still in flux.  These lifetime differences 
reflect the effects of spectator light quarks in heavy hadrons and become to 
a certain degree tractable theoretically in the limit of large mass of the 
heavy decaying quark $Q$ in terms of an expansion in the inverse powers of 
its mass $m_Q$. The leading term in this expansion is the `parton' decay 
rate $\Gamma_{part} \propto m_Q^5$ of a free heavy quark, which sets the 
overall scale for the decay rates of the heavy hadrons containing $Q$ and 
which, naturally, does not depend on the spectator quarks or antiquarks. The 
spectator-flavor dependence arises through two mechanisms$^{\rm [2 - 5]}$: 
the weak scattering (WS) and the Pauli interference (PI). The weak 
scattering corresponds to a cross-channel of the parton decay, generically 
$Q \to  q_1 \, q_2 \, \overline q_3$, where either a spectator light quark 
in a baryon scatters off the heavy quark into two light ones: $q_3 \, Q \to 
q_1 \, q_2$, or the spectator antiquark in a meson, say $\overline q_1$, 
annihilates with the $Q$ into $q_2 \, q_3$ via the $W$ boson. The Pauli 
interference effect arises when one of the final quarks in the decay of $Q$ 
is identical to a spectator light quark in the initial hadron. The 
contribution of both the WS and the PI to the decay rate of a hadron is 
suppressed with respect to the parton decay rate by the factor $m_Q^{-3}$. 
However, they come with a large numerical coefficient, typically $16 \pi^2 
/3$, reflecting mainly the difference of numerical factors in the two-body 
final phase space vs the three-body one, which makes these effects hundred 
percent significant for the charmed hadrons, as is confirmed by the 
experimental data.

A quantitative description of the spectator-flavor dependent effects in 
inclusive decay rates of charmed hadrons is problematic for two reasons:  
poor knowledge of light quark matrix elements over the hadrons (light quark 
wave functions in a naive constituent quark model) and the fact that for 
charmed hadrons the spectator effects are comparable in magnitude to the 
leading `parton' term. The latter obstacle in the way of a quantitative 
theoretical picture disappears for the $b$ hadrons, while the former one, 
the lack of understanding the hadronic matrix elements, stays. It can be 
noticed however, that an application of the calculated in the order 
$m_Q^{-3}$ WS and PI effects to lifetime differences of the charmed hadrons 
is possible at a semi-quantitative level. In particular it has allowed to 
predict$^{\cite{vs1}}$ the hierarchy of the lifetimes of the charmed mesons 
and baryons in agreement with the present data, while then existing data 
were either missing (for $\xn$), or were contradicting to the presently 
established pattern of the hierarchy (for $D_s$ and for $\Omega_c$).  
Therefore the data on the inclusive decay rates of charmed hadrons may 
provide a better information on the hadronic matrix elements, describing the 
spectator effects, than the very model-dependent theoretical 
evaluation$^{\cite{vs1}}$, thus allowing for better estimates of the 
differences of inclusive decay rates of $b$ hadrons by a rescaling of the 
heavy quark mass.

The extrapolation to the $b$ hadrons is especially interesting in view of 
the recent data$^{\cite{lep,cdf}}$, indicating a large difference of the 
lifetimes of $\Lambda_b$ and the $B$ mesons:  $\tau(B)/\tau(\Lambda_B)=1.30 
\pm 0.07$, while typical theoretical estimates for the deviation of this 
ratio from unity range from few percent$^{\cite{vs1}}$ to at most 
0.1\,$^{\cite{bigi}}$.

The purpose of the present paper is to point out that an additional 
information on the spectator quark matrix elements in the $SU(3)_{Fl}$ 
triplet of heavy baryons consisting of $\Lambda_Q$ and two $\Xi_Q$ hyperons 
can be obtained by measuring the difference of the inclusive semileptonic 
decay rates of $\Lambda_c$ and the $\Xi_c$ hyperons.  Moreover, this 
difference should be large: the rate of these decays for the $\Xi_c$ baryons 
(equal between $\xp$ and $\xn$ up to the CKM suppressed contributions) may 
exceed that for the $\Lambda_c$ by a factor of two. Besides testing this 
prediction, following from a simple extension of the analysis$^{\cite{vs1}}$ 
of non-leptonic decays, an experimental measurement of these semileptonic 
decay rates would allow to separate the relevant matrix elements for each of 
the spectator light quarks:  WS for the $d$ quark and PI for the $u$ and $s$ 
quarks.

Indeed, the lifetime difference between the baryons and the mesons is 
contributed not only by the WS and the PI but also by a flavor-independent 
effect of order $m_Q^{-2}$ related to the motion and the chromo-magnetic 
interaction of the heavy quark inside hadron$^{\cite{buv}}$ and the same 
mechanism contributes also to the difference of decay rates of baryons with 
different spin arrangement. (For a recent review see Ref. \cite{bigi}.) 
Thus for the 
triplet $(\Lambda_c,\, \Xi_c)$ the differences of the total rates provide 
only two inputs for three matrix elements relevant for WS and PI. The 
difference of the semileptonic decay rates in this triplet is determined 
only by the PI for the $s$ quark, which thus would provide the third 
experimental input.

Even stronger effect of the positive PI for the $s$ quark on semileptonic 
decay rate should be expected for the  hyperon $\Omega_c$. 
The $\Omega_c$ is known to have the fastest total decay rate in agreement 
with the theoretical prediction$^{\cite{vs1}}$.  Comparing the PI effects in 
the nonleptonic and semileptonic decays, leads to the conclusion that the 
semileptonic decay of $\Omega_c$ is enhanced approximately by the same 
amount as its nonleptonic decay rate. Thus one may expect a quite sizeable 
value of the branching ratio $B(\Omega_c \to e^+ \, \nu \, X)$, up to 
approximately 10 - 15\%.

\section{Spectator effects in nonleptonic decays}
A systematic description of the inclusive decay rates of heavy hadrons is 
performed$^{\cite{sv0,vs,vs1}}$ by applying the operator product expansion 
in powers of $m_Q^{-1}$ to the `effective Lagrangian' arising in the second 
order in the weak-interaction Lagrangian $L_W$:
\beq
L_{eff}=2 \,{\rm Im} \, \left [ i \int d^4x \, e^{iqx} \, T \left \{ L_W(x),
L_W(0) \right \} \right ]~,
\label{leff}
\eeq
in terms of which effective Lagrangian the inclusive decay rate of a heavy 
hadron $X_Q$ is given as\footnote{The nonrelativistic normalization of
heavy quark states is systematically used throughout this paper, so that
e.g. $\langle Q | Q^\dagger Q | Q \rangle =1$.}
\beq
\Gamma_X = \langle X_Q | L_{eff} | X_Q \rangle ~.
\label{gamx}
\eeq
The leading term in $L_{eff}$ gives the `parton' decay rate.
For the nonleptonic decay of the charmed quark one uses in eq.(\ref{leff}) 
the nonleptonic part of the weak Lagrangian and finds for the CKM
unsuppressed part
\beq
L_{eff, \, nl}^{(0)}= 
{G_F^2 \, m_c^5 \over 64 \pi^3} \, \eta_{nl} \, (\overline c c)~~,
\label{leffnl}
\eeq
where the factor $\eta_{nl}$ describes the QCD corrections to the decay
of free heavy quark and the CKM mixing parameters $V_{ud}$ and $V_{cs}$
are approximated by one. For the semileptonic decay rate the analogous
expression is
\beq
L_{eff,\, sl}^{(0)}=
{G_F^2 \, m_c^5 \over 192 \pi^3} \, \eta_{sl} \, (\overline c c)~~.
\label{leffsl}
\eeq
The standard `parton' prediction for the decay rate of charmed hadrons is 
obtained by applying the equation (\ref{gamx}) and taking into account that 
$\langle X_c |(\overline c c)| X_c \rangle \approx 1$ up to corrections of 
order $m_c^{-2}$. The full set of corrections of order $m_c^{-2}$ arise from 
the corrections to the latter matrix element and also from the contribution 
to the $L_{eff}$ of the dimension 5 operator $(\overline c \, \sigma_{\mu 
\nu} G_{\mu \nu} c)$. These corrections are studied in 
detail$^{\cite{buv,bigi}}$ for both nonleptonic and semileptonic decays.  
These terms split the decay rates of heavy baryons from those of heavy 
mesons and also between heavy baryons of different spin arrangement. 
However they do not 
depend on the light spectator quark flavor, and do not split the decay rates 
within an $SU(3)_{Fl}$ multiplet, i.e., say, between the $\Lambda_c$ and the 
$\Xi_c$ hyperons. In this paper we are mainly concerned with the latter 
splittings, thus in what follows we leave the $m_c^{-2}$ terms aside.

The flavor-dependent terms in the $L_{eff}$ describing the WS and the PI 
mechanisms arise in the order $m_Q^{-3}$ in the form of dimension 6 
four-quark operators. For the purpose of further discussion we reproduce 
here the complete expression$^{\cite{vs1}}$ for these terms in the effective 
Lagrangian for nonleptonic decays of charmed hadrons\footnote{This 
reiterating of the old result  may also be helpful in view of the recently 
renewed interest to the problem$^{\cite{ns,ampr}}$.} adjusted for a 
different convention about the sign of $\gamma_5$ and for a different 
normalization of the heavy quark states:
\begin{eqnarray}
&&L_{eff, \, nl}^{(3)}= {G_F^2 \, m_c^2 \over 2 \pi} \, \left \{ {1 \over
2} \, \left [ C_+^2+C_-^2 + {1 \over 3} (1 - \kappa^{1/2}) (C_+^2-C_-^2)
\right ] \, (\overline c \Gamma_\mu c)(\overline d \Gamma_\mu d) +
\right .\nonumber \\
&&{1 \over 2} \, (C_+^2-C_-^2) \, \kappa^{1/2} \, (\overline c \Gamma_\mu d)
(\overline d \Gamma_\mu c)+ {1 \over 3} \, (C_+^2 - C_-^2) \,
\kappa^{1/2} \, (\kappa^{-2/9}-1) \, (\overline c \Gamma_\mu t^a c) \,
j_\mu^a -
 \\ \nonumber
&&{1 \over 8} \, \left [ (C_++C_-)^2 + {1 \over 3} (1-\kappa^{1/2})
(5C_+^2+C_-^2-6C_+C_-) \right] (\overline  c \Gamma_\mu c + 
{2 \over 3}\overline c \gamma_\mu \gamma_5 c) (\overline u \Gamma_\mu u)
- \\ \nonumber
&&{1 \over 8} \, \kappa^{1/2} \, (5C_+^2+C_-^2-6C_+C_-)
(\overline  c_i \Gamma_\mu c_k + 
{2 \over 3}\overline c_i \gamma_\mu \gamma_5 c_k) 
(\overline u_k \Gamma_\mu u_i)-
\\ \nonumber
&&{1 \over 8} \, \left [ (C_+-C_-)^2 + {1 \over 3} (1-\kappa^{1/2})
(5C_+^2+C_-^2+6C_+C_-) \right] (\overline  c \Gamma_\mu c + 
{2 \over 3}\overline c \gamma_\mu \gamma_5 c) (\overline s \Gamma_\mu s)
- \\ \nonumber
&&{1 \over 8} \, \kappa^{1/2} \, (5C_+^2+C_-^2+6C_+C_-)
(\overline  c_i \Gamma_\mu c_k + 
{2 \over 3}\overline c_i \gamma_\mu \gamma_5 c_k) 
(\overline s_k \Gamma_\mu s_i)-
\\ \nonumber
&&\left. {1 \over 6} \, \kappa^{1/2} (\kappa^{-2/9}-1) (5C_+^2+C_-^2) 
(\overline  c \Gamma_\mu t^a c +
{2 \over 3}\overline c \gamma_\mu \gamma_5 t^a c) j_\mu^a
\right\}~.
\label{l3nl}
\end{eqnarray}
In this formula $\Gamma_\mu=\gamma_\mu \, (1-\gamma_5)$, $C_+$ and $C_-$ are 
the standard coefficients in the RG renormalization of the nonleptonic weak 
interaction from $m_W$ down to the charmed quark mass $m_c$: $C_-= 
C_+^{-2}=(\as(m_c)/\as(m_W))^{4/b}$, the powers of the parameter 
$\kappa=(\as(\mu)/\as(m_c))$ describe the `hybrid'$^{\cite{vs1,vs2}}$ 
renormalization of the operators from the scale $m_c$ down to a low 
normalization scale $\mu$, and $j_\mu^a=\overline u \gamma_\mu t^a u + 
\overline d \gamma_\mu t^a d + \overline s \gamma_\mu t^a s$ is the color 
current of the light quarks ($t^a = \lambda^a /2$ being the color 
generators). 

The terms in eq.(\ref{l3nl}) with the $SU(3)_{Fl}$ singlet operator 
$j_\mu^a$ do not produce differences of the decay rates within the same 
$SU(3)_{Fl}$ multiplet, while the rest terms containing the  $d$, $u$, and 
$s$ quarks with different coefficients give rise to those differences. For 
the charmed hyperons the term with the $d$ quark describes the 
WS\footnote{In mesons this term describes the interference of the $\overline 
d$ quark and is responsible for the difference of the lifetime of $D^\pm$ 
from those of $D^0$ and $D_s$.}, while those with the $u$ and $s$ describe 
the interference between the quark produced in the $c$ quark decay with that 
in the initial hyperon. The color antisymmetry of the baryon wave function 
relates the matrix elements over baryons of the operators with a crossed 
color arrangement:  $\langle X_c |(\overline c_i T_\mu  c_k) (\overline q_k 
\Gamma_\mu q_i) | X_c \rangle = -\langle X_c |(\overline c T_\mu  c) 
(\overline q \Gamma_\mu q) | X_c \rangle$ for any spinor structure $T_\mu$. 
Furthermore, in the leading order in $m_c^{-1}$ the axial current of the 
charmed quark is proportional to its spin, which is decoupled from that of 
the light quark pair in the spin ${1 \over 2}$ hyperons $\Lambda_c, \, 
\Xi_c$.  Thus for these baryons only the vector current of the charmed 
quarks contributes and the matrix elements of the effective Lagrangian in 
eq.(\ref{l3nl}) reduce to those of the operators $O_q=(\overline c 
\gamma_\mu  c) (\overline q \gamma_\mu q)$ with $q=u, \, d$, or $s$.  
Clearly, these matrix elements are unknown, which makes a further analysis 
model dependent. Within an additive constituent quark model applied to 
operators at a low normalization point $\mu$ the shift of the decay rates of 
the baryons due to the effective Lagrangian (\ref{l3nl}) can be written as:
\begin{eqnarray}
\Delta \Gamma_{nl} (\Lambda_c) &=& c_d  \, \langle O_d \rangle_{\Lambda_c} +
c_u \, \langle O_u \rangle_{\Lambda_c} \\ \nonumber
\Delta \Gamma_{nl} (\xp) &=& c_s \, \langle O_s \rangle_{\xp} +
c_u \, \langle O_u \rangle_{\xp} \\ \nonumber
\Delta \Gamma_{nl} (\xn) &=& c_d \, \langle O_d \rangle_{\xn} +
c_s \, \langle O_s \rangle_{\xn}
\label{difad}
\end{eqnarray}
with $\langle O_q \rangle_{X_c} = \langle X_c| O_q | X_c \rangle$ and
the coefficients $c_q(\mu)$ read off eq.(\ref{l3nl}):
\begin{eqnarray}
c_d &=& {G_F^2 \, m_c^2 \over 4 \pi} \, \left [ C_+^2+C_-^2 + {1 \over 3}
(4 \kappa^{1/2}-1) (C_-^2-C_+^2) \right ] \\ \nonumber
c_u &=& - {G_F^2 \, m_c^2 \over 16 \pi} \, \left [ (C_++C_-)^2 + {1 \over
3} (1-4 \kappa^{1/2}) (5C_+^2 + C_-^2 - 6 C_+C_-) \right ] \\ \nonumber
c_s &=& - {G_F^2 \, m_c^2 \over 16 \pi} \, \left [ (C_+-C_-)^2 + {1 \over
3} (1-4 \kappa^{1/2}) (5C_+^2 + C_-^2 + 6 C_+C_-) \right ]~.
\label{coef}
\end{eqnarray}
The numerical values of these coefficients depend on the choice of the 
normalization point $\mu$ at which the constituent quark model may be 
applied and, to a lesser extent, on the value of $\as(m_c)$. Varying $\mu$ 
in the limits such that $\as(\mu)$ varies between 0.5 and 1, one finds that 
the coefficients $c_d$ and $c_s$ are positive and $c_u$ is negative and that 
$c_d \approx - 3\, c_u$ with $c_s$ being approximately equal, or slightly 
smaller than $c_d$. If one further assumes that the matrix elements of each 
of the operators $O_q$ over the corresponding constituent quark are the 
same, one finds$^{\cite{vs1}}$ the hierarchy of the decay rates within the 
triplet of $\Lambda_c$ and $\Xi_c$ hyperons in a qualitative agreement with 
the presently observed.

The absolute value of the splitting of the rates is even more difficult
to predict, since this requires knowledge of the absolute magnitude of
the matrix elements $\langle O_q \rangle_{X_c}$. A very approximate
estimate$^{\cite{vs1}}$ in a naive nonrelativistic model indicates that
the effect of these operators on the decay rates should be comparable to
the `parton' decay rate.

\section{Spectator effects in semileptonic decays}
An additional effect, which has been disregarded in the previous studies, 
but whose experimental observation can be quite helpful in understanding the 
spectator effects in inclusive decay rates of heavy hadrons, is a similar PI 
of the $s$ quark produced in the underlying charmed quark decay $c \to s l^+ 
\nu$. The effective Lagrangian describing this interference effect can be 
readily deduced from the nonleptonic one in eq.(\ref{l3nl}). This amounts to 
taking into account the overall color factor, setting the coefficients $C_+$ 
and $C_-$ equal to one, and keeping only the contribution of the $\overline 
s \Gamma_\mu s$ operator (operators with $u$ and $d$ quarks still appear, 
through a `penguin' type renormalization, but only in terms with $j_\mu^a$). 
In this way one finds an analog of eq.(\ref{l3nl}) for the spectator effects 
in semileptonic decays:
\begin{eqnarray}
&&L_{eff, \, sl}^{(3)}= {G_F^2 \, m_c^2 \over 12 \pi} \, \left [ 
(\kappa^{1/2}-1)\, (\overline  c \Gamma_\mu c + 
{2 \over 3}\overline c \gamma_\mu \gamma_5 c) (\overline s \Gamma_\mu
s)- \right. \\ \nonumber
&&\left. 3\, \kappa^{1/2} \, (\overline  c_i \Gamma_\mu c_k + 
{2 \over 3}\overline c_i \gamma_\mu \gamma_5 c_k) 
(\overline s_k \Gamma_\mu s_i)-
\kappa^{1/2} \, (\kappa^{-2/9}-1) \,
(\overline  c \Gamma_\mu t^a c +
{2 \over 3}\overline c \gamma_\mu \gamma_5 t^a c) j_\mu^a
\right ]~.
\label{l3sl}
\end{eqnarray}

Proceeding in the same way as for the nonleptonic decays, i.e. leaving
out the $SU(3)_{Fl}$ symmetric operator containing $j_\mu^a$ and also
using the color antisymmetry of the baryon wave function, one finds the
shift of the semileptonic decay rates of the $\Xi_c$ baryons due to PI
of the $s$ quark:
\beq
\Delta \Gamma(\Xi_c \to e^+ \, \nu X) = c_{sl} \, \langle O_s
\rangle_{\Xi_c}
\label{difsl}
\eeq
with the same operator $O_s$ as in eqs.(\ref{difad}) and the coefficient
$c_{sl}$ given by
\beq
c_{sl}={G_F^2 \, m_c^2 \over 12 \pi} \, (4 \kappa^{1/2}-1)~~.
\label{csl}
\eeq
Clearly, the semileptonic decay rates of $\xp$ and $\xn$ should be equal
(up to CKM suppressed contributions) due to the isotopic symmetry.

Allowing for the variation of $\as(\mu)$ at the low normalization point
between $\as(\mu)=0.5$ and $\as(\mu)=1$, one can readily see that with a
better than 10\% accuracy the following relation holds $c_{sl} \approx
c_s /3$, where the factor of $1/3$ is clearly due to the overall color
suppression of the semileptonic decay. In other words, the relative
shift of the decay rate due to the PI of the $s$ quark is approximately
the same for semileptonic decays as for the nonleptonic ones. 

From the observed differences of lifetimes of the  charmed 
baryons and using the additive quark model relations (\ref{difad}) one can 
conservatively estimate that the effect of the constructive PI of the $s$ 
quark in these baryons enhances the nonleptonic decay rate by at least 1 
ps$^{-1}$ in absolute units.  Therefore according to the above estimate of 
the relative magnitude of the coefficients $c_{sl}$ and $c_s$ the effect of 
the PI in the semileptonic decay rates should amount to at least 0.3 
ps$^{-1}$, which is  somewhat larger or approximately equal to the measured 
semileptonic decay rate of $\Lambda_c$: $\Gamma(\Lambda_c \to e^+ \, \nu \, 
x)= 0.23 \pm 0.9$ ps$^{-1}$. Therefore we come to the conclusion that the 
semileptonic decay rate of the $\Xi_c$ baryons can exceed that of the 
$\Lambda_c$ by a factor of two, or more.

The effects of the strange quark interference are especially strong in the 
weakly decaying  baryon $\Omega_c$ and are 
considered$^{\cite{vs1}}$ to be primarily responsible for the fact that this 
hyperon has the shortest lifetime among weakly decaying charmed hadrons.  
Since these effects are approximately proportional for the semileptonic and 
the nonleptonic decay, one should expect that the semileptonic decay rate of 
the $\Omega_c$ is enhanced proportionally to the total decay rate. Thus the 
semileptonic branching ratio of $\Omega_c$ should essentially be determined 
by the usual overall color factor and the relative perturbative enhancement 
(by the coefficients $C_+$ and $C_-$) of the nonleptonic decay. Thus one 
should expect the branching ratio $B(\Omega_c \to e^+ \, \nu \, X)$ close to 
10-15\%.

\section{Summary}

The differences of the inclusive weak decay rates of charmed and beauty 
particles still present an interesting problem for a theoretical and 
experimental study. The theoretical description of these differences in 
terms of expansion in the inverse heavy quark mass runs into difficulty of 
poor knowledge of the hadronic matrix elements of the relevant operators. 
This uncertainty can be somewhat reduced by measurements for the charmed 
hadrons, where the relative differences of the rates are large. In this 
respect the measurement of the differences of semileptonic decay rates of 
charmed baryon would provide an important information on the magnitude of 
the enhancement of the decay rates due to the constructive Pauli 
interference of the $s$ quarks. According to the estimates presented in this 
paper, the differences of the absolute semileptonic decay rates can be 
large: $\Gamma(\Xi_c \to e^+ \, \nu \, X) \, 
\mbox{\raisebox{-.8ex}{$\stackrel{\textstyle >}{\sim}$}} \, 
2\,\Gamma(\Lambda_c \to e^+ \, \nu \, X)$. This interference effect should 
be further enhanced in the semileptonic decays of the $\Omega_c$, due to the 
presence of two strange quarks in the baryon. The estimates of the present 
paper suggest that the enhancement of the semileptonic decays of $\Omega_c$ 
should be approximately proportional to the overall enhancement of its total 
decay rate. Thus the semileptonic branching ratio of $\Omega_c$ can be as 
large as 10-15\%.

This work is supported, in part, by the DOE grant DE-AC02-83ER40105.


\begin{thebibliography}{99}
\bibitem{delco}
W. Bacino {\it et.al.} (DELCO Coll.),  \prl{45}{80}{329}.
\bibitem{sv0}
M.A. Shifman and M.B. Voloshin, (1981) unpublished, presented in the
review V.A.Khoze and M.A. Shifman, \spu{26}{83}{387}.
\bibitem{bgt}
N. Bilic, B. Guberina and J. Trampetic, \np{B248}{84}{261}.
\bibitem{vs}
M.A. Shifman and M.B. Voloshin, \sjnp{41}{85}{120}.
\bibitem{vs1}
M.A. Shifman and M.B. Voloshin, \spj{64}{86}{698}.
\bibitem{lep}
I.J. Kroll, {\it Masses and Lifetimes of B hadrons}, talk presented at
the 17th Int. Symp. on Lepton and Photon Interactions, Beijing, August
10-15 1995.
\bibitem{cdf}
G. Apollinari (CDF collaboration), presented at the Aspen Winter
Conference on Particle Physics, Aspen, Co., January 1996.
\bibitem{bigi}
I.I. Bigi, Univ. Notre Dame report UND-HEP-95-BIG02, June 1995;
[hep-ph/9508408].
\bibitem{buv}
I.I. Bigi, N.G. Uraltsev, and A.I. Vainshtein, \pl 
{\bf B293} 430 (1992), erratum -- ibid. {\bf B297},  477 (1993).
\bibitem{ns}
M. Neubert and C.T. Sachraida, report CERN-TH/96-19, SHEP 96-03, March
1996; [hep-ph/9603202].
\bibitem{ampr}
G. Altarelli, G. Martinelli, S. Petrarca and F. Rapuano, report
CERN-TH/96-77, ROME1 prep. 1143/96, April 1996; [hep-ph/9604202].
\bibitem{vs2}
M.A. Shifman and M.B. Voloshin, \sjnp{45}{87}{292}.

\end{thebibliography}
\end{document}